\documentstyle[12pt]{article}
\begin{document}
\baselineskip=14pt
  
\centerline{No $\Lambda$ oscillations}
   
\vspace{0.2in}
 
\centerline{J. Lowe\footnote{Also at Physics Department, University, 
Birmingham B15 2TT, England}, B. Bassalleck, H. Burkhardt\footnote{Also 
at Shell Centre for Mathematical Education, University, Nottingham NG7 2RD,
England}, A. Rusek\footnote{Present address: Brookhaven National Laboratory,
Upton, NY 11973, USA}, G.J. Stephenson Jr.}
 
\centerline{\it Physics Department, University of New Mexico, Albuquerque, 
NM 87131, USA}
  
\vspace{0.1in}
  
\centerline{and}
 
\vspace{0.1in}
 
\centerline{T. Goldman}
\centerline{\it Theoretical Division, Los Alamos National Laboratory, 
Los Alamos, NM 87545, USA}
 
\vspace{0.5in}
  
ABSTRACT: We examine a recently published calculation which predicts
an oscillatory behaviour for the decay of $\Lambda$s produced together
with a neutral kaon, and proposes a new expression for the wavelength of 
kaon strangeness oscillations. We modify the calculation by imposing
the requirement that the interference of the $K_L$ and $K_S$ components
of the kaon wave function occurs at a specific space-time point.
With this requirement, the unusual results predicted vanish, 
and the conventional results are recovered.
 
\vspace{1.0in}
  
When neutral kaons are produced in a hadronic reaction, strangeness
conservation dictates that the kaons are produced in one of
the strangeness eigenstates, either a $K^0$ or a $\bar{K}^0$. For
example, the reactions
 
$$\pi^- p\rightarrow \Lambda K^0$$
 
\noindent or
 
$$K^-p\rightarrow\bar{K}^0n$$
 
\noindent produce essentially pure $K^0$ and $\bar{K}^0$ states respectively.
These are mixtures of the well-known mass eigenstates $K_L$ and $K_S$, e.g.
 
\begin{eqnarray}
\mid K^0\rangle=\sqrt{\frac{1+\mid\epsilon\mid^2}{2(1+\epsilon)^2}}
\left(\mid K_S\rangle+\mid K_L\rangle\right)
\end{eqnarray}
 
\noindent where $\epsilon$ is the usual CP-violation parameter. Because 
the $K_L$ and $K_S$ have different lifetimes (and therefore amplitudes) 
and masses (and therefore phases), the system does not remain in a $K^0$
state, but oscillates between a $K^0$ and a $\bar{K}^0$, approaching the
equal mixture of a pure $K_L$ state. This is the well-known phenomenon of
strangeness oscillations[1]. 
 
\vspace{0.3in}
 
The dynamics of the $K^0-\bar{K}^0$ system are nearly always treated
in isolation, without regard for other particles in the process. However,
in a recent paper, Srivastava, Widom and Sassaroli[2] (denoted
by SWS in the following) examined the kinematics of the reaction
$\pi^-p\rightarrow\Lambda K^0$. They pointed out that since the $K_L$ and
$K_S$ differ in mass by $\delta m = m_{K_L} - m_{K_S} = 3.522\times 10^{-12}$
MeV, the final state contains a mixture of two momenta for both
the kaon and the $\Lambda$. In the overall center of mass, the final
state is
 
\begin{eqnarray}
\mid\Lambda K(t=0)\rangle=\sqrt\frac{1+\mid\epsilon\mid^2}{2(1+\epsilon)^2}
~~\{\mid\Lambda_SK_S\rangle+\mid\Lambda_LK_L\rangle\}
\end{eqnarray}
 
\noindent at the moment of production, $t=0$. Denoting the total 
center-of-mass energy by $\sqrt{s}$, the center-of-mass momenta of the 
$K_L$ and $K_S$ are given by
 
$$p_i=\frac{(s-m_i^2-m_{\Lambda}^2)^2-4m_i^2m_{\Lambda}^2}{4s}$$
 
\noindent where $i=L$ or $S$. $\Lambda_L$ and $\Lambda_S$ denote a 
$\Lambda$ in a momentum eigenstate with momentum equal in magnitude to that
of the corresponding 
kaon. We use the subscripts $L$ and $S$ to denote kinematic quantities 
relevant to the $K_L$ and $K_S$, quantities without a subscript to the
average of these, and the subscript $\Lambda$ for quantities relevant 
to the $\Lambda$.
A consequence of these two momenta is that there are four rest frames 
relevant to the problem, i.e. those for the $K_L$, the $K_S$, the 
$\Lambda_L$ and the $\Lambda_S$. The proper times in these frames are 
denoted $\tau_L$, $\tau_S$, $\tau_{\Lambda_L}$ and $\tau_{\Lambda_S}$
respectively. 
 
The state (2) develops in time according to
  
\begin{eqnarray}
\mid\Lambda K(t)\rangle=\sqrt\frac{1+\mid\epsilon\mid^2}{2(1+\epsilon)^2}
~~\{a_S(\tau_{\Lambda_S},\tau_S)\mid\Lambda_SK_S\rangle+
a_L(\tau_{\Lambda_L},\tau_L)\mid\Lambda_LK_L\rangle\}
\end{eqnarray}
 
\noindent where
 
\begin{eqnarray}
a_i(\tau_{\Lambda_i},\tau_i)={\rm exp}
\{-i(m_i\tau_i+m_{\Lambda}\tau_{\Lambda i})-
\frac{1}{2}(\Gamma_i\tau_i +\Gamma_{\Lambda i}\tau_{\Lambda i})\}
\end{eqnarray}
 
\noindent with $i=S$ or $L$. The four proper times, $\tau_L$, $\tau_S$, 
$\tau_{\Lambda_L}$ and $\tau_{\Lambda_S}$, are related to the time in the
overall center-of-mass frame, $t$, by the appropriate Lorentz transformations
relating a point $(\xi_i,\tau_i)$ in the frame $i$ to the point
$(x,t)$ in the overall center-of-mass frame:
  
$$\xi_i=\gamma_i(x-\beta_it)$$
$$\tau_i=\gamma_i(t-\beta_ix)$$
 
\noindent and this transformation must be chosen carefully.
 
\vspace{0.3in}
  
SWS seem to be the first to examine the relation between these proper
times. They used a prescription described in an earlier paper[3]. 
In their work, the form of the $K^0$ strangeness oscillations is
different from that given by the usual treatment and their results showed
several unconventional features, in particular:
 
\vspace{0.1in}
 
\noindent (S1) SWS derive a different relation between the wavelength
of the strangeness oscillations and the $K_L - K_S$ mass difference,
$\delta m$. Therefore, they deduce that existing measurements of
$\delta m$ from strangeness oscillations are in error by a factor $C(s)$
which is at least 2 and is much larger near threshold.
 
\vspace{0.1in}
 
\noindent (S2) The joint probability distribution, $P(x_{\Lambda},x_{K^0})=
\mid\Psi(x_{\Lambda},x_{K^0})\mid^2$ for detecting both a $\Lambda$ and a 
$K^0$ from the reaction $\pi^- p\rightarrow \Lambda K^0$ will show 
oscillations for both the $K^0$ and $\Lambda$ distributions as a function
of distance from the reaction point. For the $K^0$, these are the familiar
strangeness oscillations, but those for the $\Lambda$ are a new effect.
  
\vspace{0.1in}
 
\noindent These novel features are a consequence of their choice of
proper times and of their treatment of two distinct momenta in both the 
$\Lambda$ and kaon states.
 
\vspace{0.3in}
 
Our interest in the work of SWS was stimulated initially by the 
possibility of a direct experimental test of some of their predictions.
In the process of investigating this, we re-examined their derivation.
While this re-derivation confirmed some of their results, we found
some important differences, which changed the conclusions listed above.
This letter describes our results.
 
\vspace{0.3in}
 
The essence of the differences between our treatment and that of SWS 
lies in the relation between the four proper times in the respective
rest frames. In deriving these, our starting point is the point at which
the experimental observation is made. For either particle, say the kaon,
we choose a specific space point $x$ in the overall center-of-mass frame.
In the usual plane-wave treatment, the choice of an observation time $t$ 
is immaterial, since the wave function is present for all time; we are
free to observe at any time. However, 
in any scattering problem, it is implicitly assumed that a more realistic 
description can be obtained from the usual plane-wave treatment by
constructing wave packets. In view of this, the time of observation should
be chosen such that the wave packet (or, equivalently, the classical
particle) is present at the point $x$. In the present case, the outgoing
kaon is a superposition of two wave packets with different momenta, so
would separate after a sufficiently long time.
However, this is not an issue here, since the difference in velocity
of the $K_S$ and $K_L$ wave packets is small enough $(\sim 10^{-15})$
that they do not separate significantly before detection; it is easy to 
choose the size of the wave packets to be such that they are large compared
with the separation of their centroids over the time of the experiment 
while still small compared with the dimensions of the apparatus.
 
\vspace{0.1in}
 
Therefore, we choose the time of observation to be the average of when the 
wave packets for $K_S$ and $K_L$ (and also the classical particles) arrive
at point $x$. The mean velocity of these wave packets is
 
$$\bar{\beta}=\frac{1}{2}\left(\beta_L+\beta_S\right)$$
 
\noindent so that the point of observation in the center-of-mass frame is
 
$$(x,t)=\left(x,\frac{x}{\bar{\beta}}\right).$$
 
\noindent The choice of $\bar{\beta}$ is not at all crucial in the 
following. It could be replaced by, for example,  $\beta_L$ or $\beta_S$
without  changing any conclusions. With the choice $\bar{\beta}$, the 
proper time in the rest frame of particle $i$ is therefore
 
\begin{eqnarray}
\tau_i=\gamma_i\left(\frac{x}{\bar{\beta}}-\beta_ix\right)=
\gamma_ix\left(\frac{1}{\bar{\beta}}-\beta_i\right).
\end{eqnarray}
 
\noindent  The difference between our calculation and that of SWS can
be seen at this point. They relate the proper time $\tau_i$ to the 
space point in the center-of-mass, $x$, by
 
$$\tau_i^{SWS}=\frac{m_i}{p_i}x = \frac{1}{\beta_i\gamma_i}x$$
 
\noindent which can be expressed as
 
$$\tau_i^{SWS}=\gamma_ix(\frac{1}{\beta_i}-\beta_i)$$
 
\noindent which differs from our result by the change from
$1/\overline{\beta}$ to $1/\beta_i$. Therefore, because the velocities of
the $K_L$ and $K_S$ components differ slightly, SWS are calculating 
interference at {\it two different center-of-mass times}. Our treatment 
uses the {\it same center-of-mass} time for the detection of the 
interfering $K_L$ and $K_S$ components, hence the appearance of the same
expression for time ($x/\overline{\beta}$) in the Lorentz transformation
for both the $K_L$ and $K_S$. We require that the interference between the
two components is calculated at the {\it same} time and the {\it same} 
space point. In SWS, the {\it center-of-mass times} in their Lorentz 
transformation are $x/\beta_L$ and $x/\beta_S$ for the two components of 
the neutral kaon.
 
\vspace{0.1in}
 
It is important to realise that this difference is not the result of a
technical error in either calculation, but of a difference in principle
in the treatment of the quantum mechanics of the system. We believe that
it is an error in principle to calculate the interference between wave
functions at different points in space-time.
 
\vspace{0.1in}
 
With our expression (5) for $\tau_L$, $\tau_S$, $\tau_{\Lambda_L}$ and 
$\tau_{\Lambda_S}$, we can write the coefficients (4) as
 
\begin{eqnarray}
a_i(t)={\rm exp} \left[-i\left(m_{\Lambda}\gamma_{\Lambda i}
(\frac{1}{\overline{\beta}_{\Lambda}}-
\beta_{\Lambda i})x_{\Lambda}-m_i\gamma_i(\frac{1}
{\overline{\beta}}-\beta_i)x_K\right)\right.
\nonumber \\
-\left.\frac{1}{2}\left(\Gamma_{\Lambda}\gamma_{\Lambda i}(\frac{1}
{\overline{\beta}_{\Lambda}}-
\beta_{\Lambda i})x_{\Lambda}-\Gamma_i\gamma_i
(\frac{1}{\overline{\beta}}-\beta_i)x_K\right)\right]
\end{eqnarray}
  
\noindent and the state vector at center-of-mass time $t$ as
  
\begin{eqnarray}
\mid\Lambda K(t)\rangle=\sqrt\frac{1+\mid\epsilon\mid^2}{2(1+\epsilon)^2}
~~\{a_S(t)\mid\Lambda_SK_S\rangle+a_L(t)\mid\Lambda_LK_L\rangle\}
\end{eqnarray}
 
The interesting predictions result from selecting a specific strangeness
for the kaon. If we take the $K^0$ part of (7), we get
 
$$\Psi_{\Lambda K^0}(x_{K^0},x_{\Lambda})=\sqrt\frac{1+\mid\epsilon\mid^2}
{2(1+\epsilon)^2}~~\{a_S(t)\langle\Lambda K^0\mid\Lambda_SK_S\rangle+a_L(t)
\langle\Lambda K^0\mid\Lambda_LK_L\rangle\}$$
 
$~~~~~~~~~~~~~~~~~~~~=\frac{1}{2}\{a_S(t)+a_L(t)\}$
 
\vspace{0.1in}
 
\noindent and the $\bar{K}^0$ part has the opposite sign in the bracket,  
$\{a_S(t)-a_L(t)\}$. Writing $a_i(t)$ as $a_i(t)={\rm exp}(-ib_i-c_i)$, 
the joint
probability distribution for detection of a $K^0$ and a $\Lambda$ is
given by
 
$$P(x_{\Lambda},x_{K^0})= \frac{1}{4}\mid a_S(t)+a_L(t)\mid^2$$
 
$$~~~~~~~=\frac{1}{4}\{\mid a_S(t)\mid^2+\mid a_L(t)\mid^2+
2{\rm e}^{-(c_S+c_L)}{\rm cos}(b_L-b_S)\}$$
 
\vspace{0.1in}
 
\noindent The cosine term gives the oscillations. From (6),
 
$$b_i=m_{\Lambda}\gamma_{\Lambda i}(\frac{1}{\overline{\beta}_{\Lambda}}-
\beta_{\Lambda i})x_{\Lambda}-m_i\gamma_i(\frac{1}{\overline{\beta}}-
\beta_i)x_K.$$
 
The quantity $(b_L-b_S)$ is most readily evaluated using 
 
$$b_L-b_S=\frac{db}{dm}\delta m$$
 
\noindent together with
 
$$\frac{dp}{dm}=\frac{-m}{2p}\left[1+
\frac{m_{\Lambda}^2-m^2}{s}\right]$$
  
$$\frac{dE}{dm}=-\frac{dE_{\Lambda}}{dm}=\frac{m}{\sqrt{s}}.$$
 
\noindent From this, we obtain
 
$$\frac{db_i}{dm}=\frac{m}{p}x_K$$
 
\noindent and hence the cosine term becomes $cos(kx_K)$, where
 
\begin{eqnarray}
k=\frac{m\delta m}{p}.
\end{eqnarray}
 
\vspace{0.1in}
 
There are two striking features of this result
 
\vspace{0.05in}
 
\noindent (L1) In contrast to SWS, there is no dependence on $x_{\Lambda}$
in $db/dm$. Thus the oscillations in the $\Lambda$ probability distribution,
predicted by SWS, are not present.
Oscillations exist in the $\Lambda$ probability 
only in the sense that if we detect the $K^0$ and the $\Lambda$
at the same center-of-mass time, i.e. $x_K/\beta=x_{\Lambda}/\beta_{\Lambda}$,
then oscillations will be observed. However, these are just a consequence
of the kaon strangeness oscillations. If we choose a fixed $x_K$ (or,
alternatively, integrate over all $x_K$, corresponding to not detecting 
the $K^0$) the $\Lambda$ distribution has its usual exponentially decaying
form with no oscillation.
 
\vspace{0.05in}
  
\noindent (L2) The $K^0$ distribution oscillates with wave number
$k=m\delta m/p$, which is just the usual expression[1]. This
is perhaps surprising, since a new feature has been introduced into the
kinematics, i.e. the presence of two momenta and four proper times,
pointed out by SWS. Apparently, this does not affect the strangeness
oscillations.
 
\vspace{0.1in}
 
Our result differs from (S1) and (S2) of SWS due to a basic theoretical 
difference in the treatments. It is therefore natural to look for an 
experimental test, to cast light on the situation. The problem here is that
the calculation of SWS is applicable only in a very limited number of cases.
However, we can examine the experimental situation which was one of the
original motivations for SWS's work; they quote Fujii {\it et al.}[4], who
state that values of $\delta m$ derived from strangeness oscillation
measurements appear to be higher than those from regeneration experiments.
The paper of Fujii {\it et al.} certainly gives this impression. Further,
this would be readily explained by the result of SWS, since they find a
wavelength for strangeness oscillations that differs, for a given $\delta m$,
by over a factor of 2 from our result and hence from the conventional result.
However, more recent experiments do not confirm the assertion of Fujii 
{\it et al.}[4]. For example, Chang {\it et al.}[5] list eight measurements of 
$\delta m$ from strangeness oscillations of which only 2 early measurements
are higher than the results from regeneration experiments. 
 
\vspace{0.1in}
 
Direct measurements of strangeness oscillations have been made by many 
groups. For example, Gjesdal {\it et al.}[6] used $K_{e3}$ decays to
determine the $K^0$ and $\bar{K}^0$ components of a beam which was
initially predominantly $\bar{K}^0$. Their measurements agree well with
the conventional description, and therefore with our result. In making
this comparison, it is important to realise that Gjesdal {\it et al}.
used a value for $\delta m$ that is consistent with that from regeneration
experiments, thus confirming our expression (7) for the wavelength. The
applicability of SWS's result to ref. [6] is not clear. However, they
state[7] that the experimental conditions in the measurement on
$K^-p\rightarrow \bar{K}^0n$ by Camerini {\it et al.}[8] are such that
their theory should apply. Nevertheless, Camerini {\it et al.} measure 
a value for $\delta m$ from their experiment, $\delta m = (0.50\pm 0.15)
~\tau_S^{-1}$,  which is again consistent with the regeneration value, 
$\delta m = 0.476 ~\tau_S^{-1}$, which would seem to support our result.
 
\vspace{0.1in}
 
Several preprints have appeared recently which treat various aspects
of the problem of interest here, i.e. the quantum oscillations of a
particle produced in a 2-body final state. For example, Kayser[9]
discusses $B\bar{B}$ mixing and, in particular, the Einstein-Podolsky-Rosen
aspects. Grimus and Stockinger[10] and Goldman[11] discuss neutrino
oscillations from reactor neutrinos and pion decay respectively. Although 
related to the topic discussed here, none of these papers addresses directly
the questions dealt with in the present work. Thus no direct comparison with
these papers is possible except to note that none of them draws any
conclusion that is in conflict with our results.
  
\vspace{0.1in}
 
In summary, the paper of SWS introduces two new aspects into the 
treatment of kaon strangeness oscillation, the presence of two momenta
in the reaction that produces the kaons and a new treatment of the 
proper times of the various states. We believe that the former is correct
but without substantial effect on the experimental predictions. The latter,
which is the source of their novel predictions, seems to us in error, based
on interference between components of a wave function at at different 
space-time points. Our work incorporates just the first of these features.
To the extent that an experimental test is possible, their novel results
do not seem to be supported by experiment. We hope that a definitive
experiment to give a cleaner distinction between the two calculations will
be carried out in the near future.
 
\vspace{0.4in}
 
We are grateful to A. Widom for many communications. We acknowledge
support from the US DOE and the UK Rutherford Appleton Laboratory.
 
\vspace{0.4in}
 
\noindent {\bf References}
 
\vspace{0.1in}
 
\noindent [1] See, e.g., the treatment in W.E. Burcham and M. Jobes,
Nuclear and Particle Physics (Longmans, Essex, UK, 1995).
 
\vspace{0.1in}
 
\noindent [2] Y.N. Srivastava, A. Widom and E. Sassaroli, Phys. Lett. 
{\bf B344}, 436 (1995).
  
\vspace{0.1in}
 
\noindent [3] Y.N. Srivastava, A. Widom and E. Sassaroli, Zeitschrift
f\"{u}r Physik, {\bf C66}, 601 (1995).  
 
\vspace{0.1in}
 
\noindent [4] T. Fujii, J.V. Jovanovich, F. Turkot and G.T. Zorn, 
Phys. Rev. Lett. {\bf 13}, 253 (1964). 
 
\vspace{0.1in}
 
\noindent [5] C.Y. Chang, D. Bassano, T. Kikuchi, P. Dodd and J. Leitner,
Phys. Lett. {\bf 23}, 702 (1966).
 
\vspace{0.1in}
 
\noindent [6] S. Gjesdal, G. Presser, T. Kamae, P. Steffen, J. Steinberger,
F. Vannucci, H. Wahl, F. Eisele, H. Filthuth, V. L\"{u}th, G. Zech and
K. Kleinknecht, Phys. Lett. {\bf 52B}, 113 (1974).

\vspace{0.1in}
 
\noindent [7] A. Widom, private communication (1996).
 
\vspace{0.1in}
 
\noindent [8] U. Camerini, D. Cline, J.B. English, W. Fischbein, W.R. Fry,
J.A. Gaidos, R.D. Hantman, R.H. March and R. Stark,  Phys. Rev. {\bf 150}, 
1148 (1966).
 
\vspace{0.1in}
 
\noindent [9] B. Kayser, Proceedings of the Moriond Workshop on Electroweak
Interactions and Unified Theories, Les Arcs, France, March 1995.
 
\vspace{0.1in}
 
\noindent [10] W. Grimus and P. Stockinger, preprint hep-ph/9603430. 
 
\vspace{0.1in}
 
\noindent [11] T. Goldman, preprint hep-ph/9604357.

\end{document}